\documentclass[preprint2]{aastex}
\usepackage{graphicx}

\newcommand{\fig}[1]{Figure ~\ref{fig:#1}}
\shorttitle{Photometric redshift estimation using SVM}
\shortauthors{Wadadekar}
\begin{document}

\title{Estimating Photometric Redshifts Using Support Vector Machines}
\author{Yogesh Wadadekar}
\affil{Space Telescope Science Institute, 3700 San Martin Drive, Baltimore, MD 21218}
\email{wadadekar@stsci.edu}

\begin{abstract}

We present a new approach to obtaining photometric redshifts using a
kernel learning technique called Support Vector Machines
(SVMs). Unlike traditional spectral energy distribution fitting, this
technique requires a large and representative training set. When 
one is available, however, it is likely to produce results that are comparable
to the best obtained using template fitting and artificial neural
networks. Additional photometric parameters such as morphology, size
and surface brightness can be easily incorporated.

The technique is demonstrated using samples of galaxies from the Sloan
Digital Sky Survey Data Release 2 and the hybrid galaxy formation code
GalICS. The RMS error in redshift estimation is $<0.03$ for both
samples. The strengths and limitations of the technique are assessed.

\end{abstract}
  

\section{Introduction}

In the coming decade, ongoing and planned surveys will lead to an
exponential increase, in the quality and quantity of data available to
the astronomical community. Efficient and sensitive imaging and
spectroscopic surveys, such as the Sloan Digital Sky Survey (York et
al. 2000; hereafter SDSS), the VLT/VIRMOS survey (Le F{\` e}vre et
al. 2003), the VST survey, the Keck DEEP2 survey (Davis et al. 2003)
and several others will enable observational cosmologists to map, with
great accuracy and detail, the structure and evolution of the
universe. For effective analysis of these next generation datasets, a
wide variety of new tools will need to be developed; an accurate and
efficient redshift estimator is an important step in this effort.

In spite of the recent spectacular advances in multi-object
spectroscopy, photometric methods for redshift estimation provide the
most efficient use of telescope time for estimating redshifts of large
numbers of galaxies. There are two broad approaches to the
determination of photometric redshifts. In the {\it spectral energy
distribution (SED)} fitting technique (eg. Koo 1985; Sawicki, Lin \&
Yee 1997; Fern{\' a}ndez-Soto, Lanzetta \& Yahil 1999; Fontana et
al. 2000), a library of template spectra is used.  For redshift
determination, each template is redshifted, the appropriate extinction
correction is applied, and the resulting colors are compared with the
observed ones. Usually a $\chi^2$ fit is used to obtain the optimal
template/redshift pair for each galaxy. Such techniques are simple to
implement and computationally inexpensive on modern computers. Several
implementations are publicly available (eg. HYPERZ, Bolzonella,
Miralles \& Pell{\' o} 2000). The various techniques in this category,
vary in their choice of template SED's and in the procedure for
fitting. Template SED's may be derived from population synthesis
models (eg. Bruzual \& Charlot 1993) or are (based on) spectra of real
objects (eg. Coleman, Wu \& Weedman 1980), selected to span the range
of galaxy morphologies and luminosities. Both kinds of templates have
their failings - template SED's from population synthesis models may
include unrealistic combinations of parameters or exclude known
cases. The real galaxy templates are almost always constructed from
data on bright low redshift galaxies, and may be poor representations
of the high redshift galaxy population.

The alternative empirical best fit approach is feasible for a
data set where spectroscopic redshifts are available for a subsample of
objects. In such cases, the spectroscopic data can be used to
constrain the fit of a polynomial function mapping the photometric
data to the redshift (e.g. Connolly et al. 1995; Brunner et al. 1997;
Wang, Bahcall \& Turner 1998). The disadvantage of this approach, is
that it cannot be applied to purely photometric
datasets. Additionally, it cannot easily be extrapolated to objects fainter
than the spectroscopic limit.  This limitation is
particularly serious, because it is this regime that is often of the
highest interest.

Such techniques have the advantage
however, that they are automatically constrained by the properties of
galaxies in the real universe and require no additional assumptions
about their formation and evolution. Given the particular strengths
and weaknesses of these interpolative techniques, they are ideally
suited to exploit mixed datasets such as the VLT/VIRMOS Survey and the
Keck DEEP2 survey, which will provide spectroscopic redshifts for $>
10^5$ galaxies. The Sloan Digital Sky survey with its extensive
spectroscopy can also be exploited using such techniques.

Among the interpolative techniques, new possibilities based on
machine learning have emerged.  Firth, Lahav \& Somerville (2003),
Tagliaferri et al. (2002), Vanzella et al. (2004) and Collister \&
Lahav (2004) propose methods to estimate the photometric redshift
using artificial neural networks (ANNs). The levels of accuracy
achievable for photometric redshifts with ANNs is comparable to --if
not better than-- that achievable with SED fitting, in
cases where moderately large training sets are
available. Nevertheless, neural networks have some
disadvantages. Their architecture has to be determined {\em a priori}
or modified during training by some heuristic. This may not
necessarily result in the most optimal architecture. Also, neural
networks can get stuck in local minima during the training stage. The
number of weights depends on the number of layers and the number of
nodes in each layer. As the number of layers and/or nodes increase,
the training time also increases.

In this paper, we propose the use of a method from a distinct class of
machine learning methods known as the kernel learning methods. The
method --called Support Vector Machines-- is (like the ANNs) only
applicable to ``mixed'' datasets where a moderately large training
set, with photometry in the survey filters and spectroscopic redshifts
for the same objects, is available.

This paper is organized as follows. In Section 2 we provide a brief
overview of Support Vector Machines (SVMs) and an introduction to
aspects relevant to this work. In Sections 3 and 4 we apply the
technique to data from the SDSS Data Release 2 and GalICS
simulations. Section 5 discusses the results, and the regime of
applicability of this technique.

\section{Support Vector Machine (SVMs)}

Support Vector Machines (SVMs) are learning systems, that use a
hypothesis space of linear functions in a high dimensional feature
space, trained with learning algorithms from optimization theory, which
implements a learning bias derived from statistical learning
theory. 

The input parameters (eg. broadband colors) form a set of orthogonal
vectors that define a hyperspace. Each galaxy would then represent a
point in this hyperspace. The basic SVM is a linear classifier; 
i.e. training examples labeled either "yes" or "no" (eg. the answer
to the question, Is $z>1$?) are given and a maximum-margin hyperplane splits the
"yes" and "no" training examples in the hyperspace, such that the
distance from the closest examples to the hyperplane (the margin) is
maximized. The use of the maximum-margin hyperplane is motivated by
statistical learning theory, which provides a probabilistic test error
bound that  is minimized when the margin is maximized. The parameters
of the maximum-margin hyperplane are derived by solving a quadratic
programming (QP) optimization problem. There exist several specialized
algorithms for quickly solving the QP problem, that arises in SVMs.

The original optimal hyperplane algorithm was a linear classifier, and thus inapplicable to non-linear
problems.  Vapnik (1995) suggested applying
Mercer's theorem to the problem of finding maximum-margin
hyperplanes. The theorem states that any positive semi-definite kernel
function can be expressed as a dot product in a high-dimensional
feature space. The resulting algorithm is formally similar, except
that every dot product (the distance measure) in the feature space is
replaced by a non-linear kernel function operating on the input
space. In this way, non-linear classifiers can be created. The
dimensionality of the feature space depends upon the kernel function
used, eg. if the kernel used is a radial basis function, the
corresponding feature space is a Hilbert space of infinite
dimension. Maximum margin classifiers are well regularized, so the
infinite dimension does not affect the results.

An additional complication in real data could be that no
hyperplane exists that can cleanly split the "yes" and "no" examples. For
such situations, a modified maximum margin idea was introduced; the
Soft Margin method chooses a hyperplane that splits the examples as
cleanly as possible, while still maximizing the distance to the
nearest cleanly split examples.

The estimation of a continuous output parameter such as redshift requires
the extension of the SVM algorithm to handle regression. The support
vector regression algorithm was proposed by Smola (1996). The model
produced by Support Vector Classification, as described above, only
depends on a subset of the training data, because the cost function
for building the model, does not care about training points that lie
beyond the margin. Analogously, the model produced by Support Vector
Regression only depends on a subset of the training data, because the
cost function for building the model ignores any training data that is
close (within a threshold $\epsilon$) to the model prediction.

In the last decade, extensive enhancements to all aspects of the SVM
formulation have been introduced, and the field continues to be
extremely active in theoretical developments. SVMs are being used in a
wide variety of applications such as text recognition, face detection,
weather forecasting, financial market predictions and gene data
analysis. In astronomy, SVMs have recently been applied for
classifying variable stars (Wozniak et al. 2001), determining galaxy
morphology (Humphreys et al. 2001), and distinguishing AGN from stars
and galaxies (Zhang \& Zhao 2003).

Unlike ANNs, SVMs do not require choice of an
architecture before training. Any number of input dimensions can be
treated. For a detailed explanation of the mathematical underpinnings
of SVMs see Vapnik (1995,1998). For a more practical introduction,
see Cristianini \& Shawe-Taylor (2000).

An important advantage of the SVM is that adding additional input
parameters to the classifier, leads only to near linear increase in
computational cost (Smola 1996). This is an advantage for the problem
of estimating photometric redshifts because one then has the potential
to use additional photometric parameters that may be be related to
redshift (albeit in a very non-linear manner). Such parameters, which
may include size measures such as Petrosian radii or scale radii from
de Vaucouleurs or exponential fits, central surface brightness, fixed
aperture magnitudes are available in many modern survey catalogs, such
as the SDSS catalog. On the other hand, parameters such as the
disk-to-bulge luminosity ratio, which show only a weak dependence on
redshift, if any, are not likely to be useful in improving the
accuracy of redshift estimation.

Several robust software implementations of the SVM algorithm are
publicly available and each has its own set of distinguishing features
making it optimal for a particular class of classification or
regression problem. After a survey of the capabilities of the
available implementations, we decided to use the SVMTorch II program
(Collobert \& Bengio 2001) for this work. SVMTorch is a C++
implementation that works for both classification and regression
problems. It has been specially tailored to large scale problems (such
as more than 20000 examples, even for input dimensions higher than
100). A special feature of SVMTorch is that it employs a RAM-based cache to
store  the values of the most used variables of
the kernel matrix. The size of the cache that SVMTorch should use
needs to be set by the user, depending on the free memory
available.\footnote{The SVMTorch software is freely available for
academic use from http://www.idiap.ch/learning/SVMTorch.html}

For using SVMs, the user needs to choose the kernel to be used for the
mapping from the non-linear space (containing the data) to the feature
space, where the minimization is performed, using linear learning
machines. Choosing a kernel for the SVM is analogous to choosing the
architecture of the ANN. As with ANNs there is no simple heuristic for
making this choice; some experimentation with the input data is
essential. Commonly used kernel functions include polynomial, gaussian
and sigmoidal forms. If these prove ineffective, more elaborate
kernels can be utilized. For input data, that are not strongly
clustered with respect to the output variable, at least one of the
above mentioned kernels should work well. Some experimentation reveals
that the gaussian kernel with a sigma of 1.0 gives the best results
for our problem. All the results presented in this paper were obtained
using a gaussian kernel.

The second parameter that must be set in SVMTorch is the size of the
error pipe. This basically influences the number of training iterations
that are performed before the training is considered
complete. Decreasing the width of the error pipe (eps) beyond a point
increases training time considerably, with only a marginal improvement
in the final results. Through trial and error, we determined that a value
of $eps = 0.02 $ was appropriate. Unless otherwise noted, this error
pipe width has been used throughout this paper.

The key components of the system are in place once the choice of kernel and termination criterion has been made. Unlike some other learning
systems, SVMs do not require a lengthy series of experiments in which
various parameters are tweaked, until satisfactory performance is
achieved. In many cases, the most straightforward SVM implementations
are known to perform as well as other learning techniques, without any
need for further adaptation.

With this set of parameters, training a set of 10000 objects with
five input vectors requires about 2 million iterations that take about 20
minutes on an Athlon XP 1800+ processor with 512 MB of memory.

\section{Photometric redshift from SDSS data}

The Sloan Digital Sky Survey consortium have publicly released more
than $10^5$ spectroscopic redshifts for galaxies in the Data Release 2
(DR2). In order to build the training and test sets, we first selected
from the current version of the SDSS catalog database (BESTDR2) all
objects satisfying the following criteria: (1) the spectroscopic
redshift confidence must be greater than 0.95 and there must be no
redshift warning flags and (2) $0.01 < z < 0.5$ and (3) $r <
17.5$. These criteria resulted in a galaxy sample of 139 000. The
order of the catalog was randomized, and non-overlapping training and
testing sets of equal size (10,000 objects) were selected. The input vectors were the
dereddened magnitudes in each of the 5 SDSS filters, and the output
vector was the redshift.

\begin{figure*}
\epsscale{1.3}
\plotone{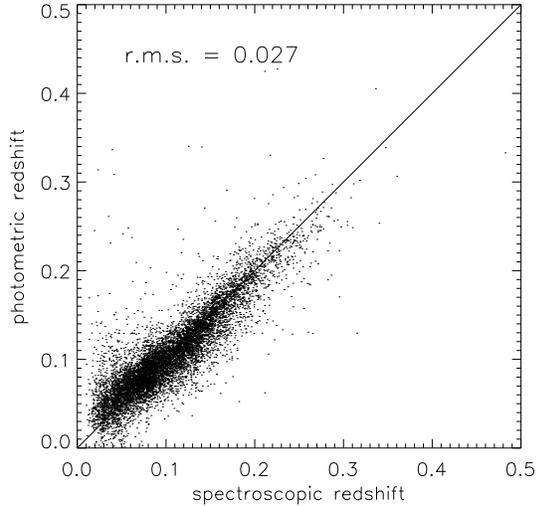} 
\caption{A comparison of photometric and spectroscopic redshifts using SDSS DR2 data. A training test of 10000 objects was used. The SVMs were tested using a non-overlapping set of 10000 objects (plotted). The rms shown is that of $\Delta_z=z_{\rm spec} - z_{\rm phot}$}\label{fig:sdssz}
\end{figure*}

\fig{sdssz} plots the SVM estimated photometric redshift against the
spectroscopic redshift for each galaxy in the test set. The rms
deviation $\sigma_\mathrm{rms} = \sqrt{\langle{}(z_\mathrm{phot} -
z_\mathrm{spec})^{2}\rangle{}} = 0.027$. The mean deviation is
0.0006. The number of outliers is small and there are no obvious
systematic deviations. There is some tendency for redshift to be
overestimated by the SVM for $z < 0.05$. Also, the scatter is larger
for $z > 0.25$. Both of these effects are caused by the presence of
fewer training examples in the sample at very low z (because the cosmic
volume sampled is small) and at higher z where incompleteness sets in.

For galaxies with $z < 0.05$, the relative fraction of training
examples can be increased substantially by constructing a separate
training set that is uniformly sampled in redshift. From our galaxy
sample of 139 000 objects, we constructed non-overlapping training and
test sets, with the additional constraint that there should be equal
number of galaxies in each redshift bin of width 0.025. Our new
training set had 7310 examples and the test set had 7300 examples. The
rms deviation was 0.029 which is somewhat larger than the
non-uniformly sampled case. This is presumably because we have fewer
galaxies in our training sample. However, when we compared only those
galaxies with $z < 0.05$, we found that the rms deviation decreased
from 0.031 to 0.029 going from the non-uniformly sampled training/test sets
to the uniformly sampled ones.

In Table 1, we compare the rms value obtained using SVMs, with that
obtained with a variety of other SED template and empirical best fit
techniques by Csabai et al. (2003) and Collister \& Lahav (2004) on a
similar sample of galaxies drawn from the SDSS Early Data Release.  The
SVM approach is clearly better than the template fitting techniques
(CWW, Bruzual-Charlot) and is nearly as good as the best empirical
best fit approach (ANN$z$).

\begin{deluxetable}{lc}

\tablecaption{Comparison of Photometric redshift accuracies using conventional template fitting, ANNz and SVM}

\tablehead{\colhead{Estimation Method} & \colhead{$\sigma_\mathrm{rms}$}}

\startdata
CWW & 0.0666 \\
Bruzual-Charlot & 0.0552 \\
Interpolated & 0.0451 \\
Polynomial & 0.0318 \\
Kd-tree & 0.0254 \\
ANN$z$ & 0.0229\\
SVM & 0.027\\
\enddata

\tablecomments{The first five entries are the photometric redshift accuracies obtained by Csabai et. al (2003) for a sample from the SDSS Early Data Release. The sixth entry is  the accuracy reached by ANN$z$, a ANN based approach (Collister \& Lahav 2004). The SVM performance was evaluated for a sample drawn from the SDSS DR2.}
\end{deluxetable}
 
\subsection{Using Additional input parameters}

One advantage of the empirical best fit approach to photometric
redshift estimation is that additional parameters, that may help in
estimating the redshift, can be easily incorporated as additional input
columns. However, these parameters need to be chosen carefully such
that they have a genuine dependence on the redshift. We found that choice of
inappropriate parameters that have no obvious redshift dependence
(eg. galaxy ellipticity) leads to larger scatter in
redshift estimation.

To illustrate this capability, we included the $r$-band 50\% and 90\%
Petrosian flux radii of our SDSS training sample as additional inputs
to the SVM. These are the angular radii containing the stated fraction
of the Petrosian flux. Each of these radii is a measure of the angular
size of the galaxy, which is a redshift dependent property. Their
ratio defines the ``concentration index'' of the galaxy, which is a
measure of the steepness of its light profile. The index is (weakly)
correlated with galaxy morphology (and therefore color).

The SVM was retrained with 7 input parameters (the 5 filter magnitudes
and the 2 Petrosian radii) for the 10000 galaxies, in our original training
sample.  When tested on the test set, the SVM produced a redshift
estimate with a rms of $\sigma_{\rm rms}=0.023$. This represents
nearly 15\% improvement in accuracy of redshift estimation, and
illustrates how additional parameters can be incorporated with little
 effort.

\subsection{Using smaller training sets}

The SDSS includes spectroscopic follow-up, of a substantial number of
galaxies detected by its photometric component. Most other surveys
lack the necessary resources, for such extensive spectroscopic
followup. In such situations, a large sample for SVM training
will not be available. We therefore need to explore the effectiveness
of the SVM on smaller training sets.

We constructed training and test sets that were 1/10 and 1/100 the
size of our original sample sets of 10 000. When the SVM was run on
these smaller data sets the rms error was respectively $\sigma_{\rm
rms}=0.036$ and $\sigma_{\rm rms}=0.049$. Clearly 100 training
examples chosen at random are insufficient to encapsulate the
diversity of the SDSS; even 1000 galaxies are not quite good
enough. This deterioration of SVM performance with smaller sample sizes
is somewhat more severe than that observed by Collister \& Lahav
(2004) with their ANN based approach.
 
\subsection{Spectral Class from SDSS data}

SED template matching techniques provide useful supplementary
information by assigning a spectral type to the galaxy, based on the
type of the best fit galaxy SED. Firth, Lahav \& Somerville (2003)
and Collister \& Lahav (2004) have demonstrated how ANNs may be used
to determine galaxy spectral type from broadband photometry.

\begin{figure*}
\epsscale{1.3}
\plotone{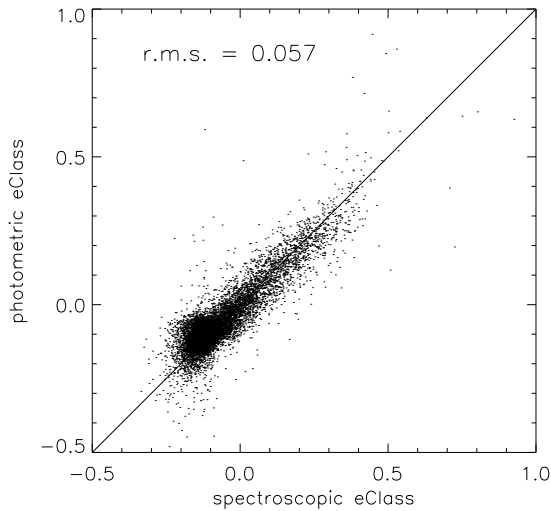} 
\caption{Results from using SVM to predict the spectral type, as measured by the {\it eClass} parameter for 10,000 galaxies from the SDSS DR2}\label{fig:sdsssc}
\end{figure*}

The spectroscopic catalog of the SDSS includes a continuous parameter
($eClass$) indicating a spectral type, deduced from an analysis of the
galaxy spectrum that ranges from about -0.5 (early type galaxies) to 1
(late type galaxies). We trained the SVM with the same 10,000 galaxies
used for redshift estimation, using $eClass$ as the output parameter,
in place of the redshift. When tested with the original test sample,
the eClass was estimated with a rms error of $\sigma_{\rm rms} =
0.057$ (\fig{sdsssc}). The error is comparable to that obtained by
Collister \& Lahav (2004) on their sample of SDSS galaxies
$\sigma_{\rm rms} = 0.052$.

\section{Photometric redshift from GalICS simulations}

One of the limitations of the SDSS survey is that it contains very few
galaxies with redshift $z >0.4$ or so. Training (and test) sets of
adequate size beyond this redshift will be difficult to obtain even
when the survey is completed. In order to test the performance of 
SVMs beyond this redshift, one needs to use galaxy magnitudes computed
from simulated models of the high redshift Universe.

To generate a mock galaxy catalog in which to train and test SVMs,
a hybrid numerical and semi-analytic model GalICS (Galaxies in
Cosmological Simulations) was used. In the model, dark matter
evolution is traced using numerical simulations, and galaxy formation
within the dark matter haloes, is treated using semi-analytic
recipes. These recipes attempt to parameterize galaxy formation, within
the framework of the hierarchical paradigm of galaxy formation. Once
the distribution and physical properties of the baryons have been
determined, a set of models are used to calculate the amount of light
they produce. Luminosities at different wavelengths are calculated
from stellar synthesis models that take into account the metallicity
and age of the stellar population. Geometry and metallicity dependent
models for the absorption and re-emission of starlight, by the dust and
gas in the interstellar medium, are included. Various lines of sight
through the simulation box generate mock galaxy catalogs that have a
realistic distribution of galaxy types, luminosities, colors and
redshift. The GalICS model is described by Hatton et al. (2003). The
mock galaxy catalog generation process is described by Blaizot et
al. (2005).

From the {\tt galics1} database of GalICS (accessible through the Mock
Map Facility\footnote{See http://galics.iap.fr} [MoMaF]), we selected 6965
objects brighter than $r_{\rm AB}=21.5$ distributed over one square
degree of sky. At this magnitude limit, GalICS is nearly free of
incompleteness introduced by its limited mass resolution at high
redshifts. At the same time, the limit is faint enough to allow us to
obtain reasonably large training and test sets, out to a redshift of $z
\sim 1$. As with the real SDSS data, the input parameters were just
the $ugriz$ magnitudes in the 5 Sloan filters and the output parameter
was the redshift. Our training and test sets were constructed by
splitting the GalICS sample into 3483 and 3482 objects respectively.
The training was done with a error pipe width $eps=0.01$. The somewhat
lower choice of the $eps$ parameter was motivated by the absence of
photometric noise in the mock galaxy catalog.

\begin{figure*}
\epsscale{1.3}
\plotone{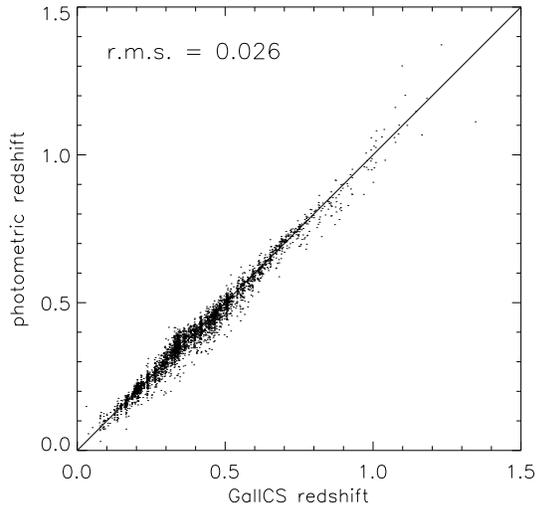} 
\caption{Photometric redshift versus GalICS model redshifts using 3482 test set galaxies. A non-overlapping set of 3483 galaxies was used for training.}\label{fig:galicsz}
\end{figure*}

\fig{galicsz} plots the SVM estimated photometric redshift for the test
sample against the redshift in the GalICS model. The rms
scatter is $\sigma_\mathrm{rms}=0.026$. There are no systematic
deviations and virtually no outliers. The complete lack of outliers is
probably due to the fact that GalICS spectra are based on a restricted
set of model templates.

\section{Discussion}

The SVM technique presented here provides comparable performance to
ANN based techniques.  The major advantage over ANNs is that the SVMs
require lesser effort in training, eg. in using ANNs the researcher
has to make a decision about the optimal network architecture (number
of layers, number of nodes in each layer, committee of networks
required to minimize network variance etc.). More complex network
architectures have more free parameters (weights) and therefore allow
a closer fit to the data, but are subject to the danger of
overfitting.  Also, adding layers or nodes to the network, leads to an
increase in training time. The challenge for the ANN expert is to find
the simplest possible architecture that provides satisfactory
results. SVM simplifies this process by replacing the ``choice of
architecture'' problem with one of ``choice of kernel'' (and
associated kernel function parameters). As we have seen, even simple
kernel functions such as a Gaussian, give a performance comparable to
that obtained with finely tuned ANNs.

Like ANNs, the technique is well suited only for cases where the
training set is available with the same general characteristics (of
magnitude and color distributions) as the sample for which photometric
redshifts are to be determined. This means that extrapolation beyond
the spectroscopic limit is not straightforward, except in situations
where the color characteristics of the fainter population are
independently constrained eg. Collister \& Lahav (2004) show how
photometric redshifts may be obtained for faint Luminous Red Galaxies,
that are about a magnitude fainter than the spectroscopic limit, using
a neural network trained on their brighter, lower redshift
counterparts. In this case, some extrapolation is possible, because
these early-type galaxies show little spectral evolution with
redshift. For a less constrained galaxy sample, such extrapolation is
not possible. Operationally, this implies that spectroscopic followup
that covers a limited area of sky but is very deep, is preferable to a
followup program that is shallow, but covers a wider area. Of course,
this may not be compatible with other aims of the survey.  Secondly,
the size of the training set has to be sufficiently large. To obtain
redshifts in the range $z=0$ to $z=3$ we estimate that about $10^4$
galaxies in the training set will be needed with photometric data in 5
optical bands. It may be possible to reduce the size of the training
set if broadband photometry in additional filters is
available. Ideally, the galaxies in the training set should be
properly distributed in redshift bins. This never occurs in a flux
limited survey. As a result, results at higher $z$ where the training
examples are less numerous will have larger errors in redshift
estimation. At high redshifts, a combination scheme of independent
photometric redshift estimation through SED fitting and SVM/ANN
techniques, may help restrict outliers to genuinely astrophysically
distinct galaxies.

It must be noted that the rms errors on redshift estimation reported
in this paper, only apply to the test set as a whole. Error bars on
individual galaxy redshifts are not available. So, although the
redshifts are likely to be estimated correctly on the average, the
redshift of a particular galaxy may be off by a larger amount. This
shortcoming applies to most {\it empirical best fit} type approaches.
These techniques are thus not well suited, to finding rare objects in the
test set, that do not have numerous corresponding examples in the training
set. On the other hand, it is well suited to problems that require the
redshift distribution rather than accurate redshifts of individual
galaxies eg. for mapping the large scale structure.

In principle, it is possible to train SVMs to the depth desired using
simulated catalogs --from models such as GalICS-- and then apply the
trained SVMs, for photometric redshift estimation from real data. Such
an approach has been taken using ANNs by Vanzella et al. (2004). We have
not attempted the same with the GalICS simulations, as the current
version is seriously incomplete at high z and low luminosities. Also
the simulation box is too small, to provide a large enough sample size
for training. A larger simulation with more particles and a larger box
size ({\tt galics3}) is under process with GalICS.

Deep spectroscopic surveys such as DEEP2 and VLT/VIRMOS will provide
the large training sets necessary for SVM applications. With the high
quality of data obtained by these surveys, the level of photometric
redshift estimation should be quite accurate to a redshift of $z\sim 1$.

SVMs may well be appropriate to other regression and classification
problems in astronomy.

\acknowledgements

This work uses the GalICS/MoMaF Database of Galaxies.\footnote{See http://galics.iap.fr}

Funding for the creation and distribution of the SDSS Archive\footnote{The SDSS website is http://www.sdss.org.} has been
provided by the Alfred P. Sloan Foundation, the Participating
Institutions, the National Aeronautics and Space Administration, the
National Science Foundation, the U.S. Department of Energy, the
Japanese Monbukagakusho, and the Max Planck Society.

The SDSS is managed by the Astrophysical Research Consortium (ARC) for
the Participating Institutions. The Participating Institutions are The
University of Chicago, Fermilab, the Institute for Advanced Study, the
Japan Participation Group, The Johns Hopkins University, Los Alamos
National Laboratory, the Max-Planck-Institute for Astronomy (MPIA),
the Max-Planck-Institute for Astrophysics (MPA), New Mexico State
University, University of Pittsburgh, Princeton University, the United
States Naval Observatory, and the University of Washington.

The author thanks the anonymous referee whose insightful comments helped improve this paper. The author also thanks N. S. Philip for bringing the SVM to his
attention. This research was initially supported by projects 1610-1
and 1910-1 of the Indo-French Center for the Promotion of Advanced
Research (CEFIPRA). Support for program AR 9540 was provided by NASA
through a grant from the Space Telescope Science Institute, which is
operated by the Association of Universities for Research in Astronomy,
Inc., under NASA contract NAS 5-26555.

\end{document}